# Estimation of the information contained in the visible matter of the universe

Melvin M. Vopson[1]

[1] School of Mathematics and Physics, University of Portsmouth, PO1 3QL Portsmouth, UK
E-mail: melvin.vopson@port.ac.uk

**Abstract**

The information capacity of the universe has been a topic of great debate since 1970s, and continues to stimulate multiple branches of physics research. Here we used Shannon's information theory to estimate the amount of encoded information in all the visible matter in the universe. We achieved this by deriving a detailed formula estimating the total number of particles in the observable universe, known as the Eddington number, and by estimating the amount of information stored by each particle about itself. We determined that each particle in the observable universe contains 1.509 bits of information and there are $\sim 6 \times 10^{80}$ bits of information stored in all the matter particles of the observable universe.

## 1. Introduction

The issue of information content associated with the physical measurement, the matter and, by extension, the information content of the universe, has been a topic of scientific debate since late 1920s. Leo Szilard analysed the relationship of information to physical process in 1929 [1], demonstrating that information about a system dictates its possible ways of evolution and behavior, and offering an elegant solution to Maxwell's famous paradox, known as the Maxwell's Demon [2].

With the emergence of digital computers, digital technologies and digital data storage, the topic of information physics entered a new era. In 1961, Rolf Landauer first proposed a link between thermodynamics and information, demonstrating that digital information is not just an abstract mathematical entity, but it is in fact physical [3]. Later, Landauer speculated that the universe could be a giant computer simulating itself [4], an idea further developed in more recent studies by Lloyd [5-7]. Wheeler considered the universe made up of particles, fields and information, proposing that everything emanates from the information inherent within it, i.e. "It from Bit" [8]. The Landauer's principle and the physical nature of information has been recently demonstrated [9-12], and it also stimulated several new studies and novel concepts including the Black Hole information paradox and the holographic principle [13], the emergent gravity [14,15] and the quantized inertia [16,17]. Other studies extrapolated the physical nature of information to estimate the mass of information [18-21], culminating with the recent publication of the mass-energy-information equivalence principle [22], and Vopson's postulate that "information" is the 5$^{th}$ form of dominant matter in the universe along solid, liquid, gas and plasma [23, 24].

These radical theories are based on the principle that information is physical, the information is registered by physical systems, and all physical systems can register information [25]. Accordingly, there is a given amount of information stored in the universe, regardless whether it is observed or not. The proposed existence of this information imposes some fundamental questions about it: "*Why is there information stored in the universe and where is it?*", and "*How much information is stored in the universe?*". Let us deal with these questions in detail.

To answer the first question, let us imagine an observer tracking and analyzing a random elementary particle. Let us assume that this particle is a free electron moving in the vacuum of space, but the observer has no prior knowledge of the particle and its properties. Upon tracking the particle and commencing the studies, the observer will determine, via meticulous measurements, that the particle has a mass of $9.109 \times 10^{-31}$ kg, charge of $-1.602 \times 10^{-19}$ C and a spin of *1/2*. If the examined particle was already known or theoretically predicted, then the observer would be able to match its properties to an electron, in this case, and to confirm that what was observed / detected was indeed an electron. The key aspect here is the fact that by undertaking the observations and performing the measurements, the observer did not create any information. The three degrees of freedom that describe the electron, any electron anywhere in the universe, or any elementary particle, were already embedded somewhere, most likely in the particle itself. This is equivalent to saying that particles and elementary particles store information about themselves, or by extrapolation, there is an information content stored in the matter of the universe. Due to the mass-energy-information equivalence principle [22], we postulate that information can only be stored in particles that are stable and have a non-zero rest mass, while interaction / force carrier bosons can only transfer information via waveform. Hence, in this work we are only examining the information content stored in the matter particles that make up the observable universe, but it is important to mention that information could also be stored in other forms, including on the surface of the space-time fabric itself, according to the holographic principle [13].

The second question is: *"How much information content is there in the observable universe?"*

This question has been addressed in several studies going back as far as late 1970s. Using Bekenstein – Hawking formula for the black-hole entropy [26, 27], Davies calculated the information content of the universe by applying the black-hole entropy formula to the whole universe [28]:

$$I \approx \frac{2\pi G M_u^2}{hc} = 10^{120} \, bits \qquad (1)$$

where *G* is the gravitational constant, $M_u$ is the mass of the universe enclosed within its horizon, *h* is Planck's constant and *c* is the speed of light. In fact, Davies produced a time dependent information formula, that allows studying the time evolution of the information in the universe, as discussed in detail by Treumann [29].

Wheeler estimated the number of bits in the present universe at *T = 2.735 K* from entropy considerations, resulting in $8 \times 10^{88}$ bits content [8]. Lloyd took a similar approach and estimated the total information capacity of the universe as [6]:

$$I = \frac{S}{k_B \ln(2)} \approx \left(\rho c^5 t^4 / \hbar\right)^{3/4} \approx 10^{90} \text{ bits} \qquad (2)$$

where *S* is the total entropy of the matter dominated universe, $k_B$ is the Boltzmann constant, $\rho$ is the matter density of the universe, *t* is the age of the universe at the present time, *c* is the speed of light, and $\hbar$ is the reduced Planck constant.

Applying Landauer's principle, Gough developed an information equation of state and calculated that $10^{87}$ bits of intrinsic universe information content could account for all the Dark Energy [30]. His estimates are for a cosmic time coinciding with stars formation, when most of the universe bit information content is represented by high temperature baryons giving an average bit energy value of 120 eV [30]. Using the mass-energy-information equivalence principle [22], Vopson estimated that around $52 \times 10^{93}$ bits would be enough to account for all the missing Dark Matter in the observable universe [22,24]. However, this is most likely overestimated, as he assumed that all the information bits are stored at *T = 2.735 K*. This is inaccurate because a significant amount of the baryonic matter is contained in stars, intergalactic gas and dust, which all have temperatures larger than the CMB.

The prime objective of this study is to estimate the bits of information contained in the matter particles of the observable universe via a new approach, using Shannon's information theory. This approach is therefore independent of the dynamics of the universe expansion, or the internal dynamics in the universe, giving an accurate bit content at any given time of observation.

## 2. The information content per elementary particle

The amount of information extracted from observing the occurrence of an event can be estimated using Shannon's information theory formalism published in 1948 [31]. Ignoring any particular features of the event, the observer or the observation method, Shannon developed his theory using an axiomatic approach in which he defined information *(I)* extracted from observing an event as a function of the probability *(p)* of the event to occur or not, *I(p)*, and he identified that the only function satisfying the imposed axiomatic properties is a logarithmic function. Hence, for a single event whose probability of occurring is *p*, the information extracted from observing the event is:

$$I(p) = -\log_b p = \log_b(1/p) \qquad (3)$$

where *b* is an arbitrary base, which gives the units of information, i.e. for binary bits of information, *b = 2*. Let us assume a set of *n* independent and distinctive events $X = \{x_1, x_2, \ldots, x_n\}$ having a probability distribution $P = \{p_1, p_2, \ldots, p_n\}$ on *X*, so that each event $x_j$ has a probability of occurring $p_j = p(x_j)$, where $p_j \geq 0$ and $\sum_{j=1}^{n} p_j = 1$. According to Shannon, the average information per event, or the number of bits of information per event, one can extract when observing the set *X* once is:

$$H(X) = -\sum_{j=1}^{n} p_j \cdot \log_b p_j \qquad (4)$$

*H(X)* is called the information entropy function and it is maximum when the events $x_j$ have equal probabilities, $p_j = 1/n$, so $H_{max}(x) = \log_b n$. When observing *N* sets of events *X*, or equivalently observing *N* times the set of events *X*, the number of bits of information extracted from the observation is *N·H(X)*. To exemplify this, let's assume a two-state system corresponding to a digital bit 0 or 1. Hence, *n = 2* and *X = {0,1}*, and assuming no biasing or external work on the system, then the two states have equal probabilities, $p_j = 1/n = 1/2$ and $p = \{p_1, p_2\} = \{1/2, 1/2\}$. Using (4), the information entropy of this system is:

$$H(X) = -\left(\frac{1}{2}\log_2\left(\frac{1}{2}\right) + \frac{1}{2}\log_2\left(\frac{1}{2}\right)\right) = \log_2 2 = 1 \quad (5)$$

*H(X) = 1* means that 1 bit of information is encoded in this process, or conversely, observing the above event generates 1 bit of information.

We will use the same approach to estimate the information content of matter in the universe. The observable universe is made up of building blocks called elementary particles. According to the Standard Model, there are 30 elementary particles and antiparticles known today. These are six quarks and their corresponding anti-particles, six leptons and their corresponding anti-particles, five vector bosons and one scalar boson. Out of all these particles, only a fraction of them are participants of the observable universe, as observed today. Most of the other particles or anti-particles are unstable and have lifetimes extremely short, so their observation is only possible via artificially created experimental conditions or theoretically. Therefore, their participation to the observable universe is negligible and, by extrapolation, their capacity to register information is also negligible. As already postulated in the previous section, all force carrier particles / bosons and particles with zero rest mass cannot store / register information and they can only transfer information, so they are not considered in our estimates.

Each elementary particle is fully described by at least three degrees of freedom: mass, charge and spin. As we already discussed, these degrees of freedom are embedded within each elementary particle, like an intrinsic label, or "particle DNA". We regard this as the information content registered in each particle about itself, measured in bits. Excluding the bosons, the anti-particles and all the unstable elementary particles, out of all 30 elementary particles, only protons, neutrons and electrons are actual participants to the observable universe today. Of course, protons and neutrons are in fact not elementary particles, as they are made up of three quarks each, which would need to be accounted for. We begin however our estimate accounting only for protons, neutrons and electrons, and later we introduce the quarks.

The key aspect is to regard each elementary particle in the universe as an event in Shannon's information theory formalism. In order to calculate the information content per elementary particle in the observable universe, we use the percentage weight abundance of elements in the universe, which are about 75% Hydrogen, 23% Helium and 2% heavy elements (A > 4) [see the Supplementary Material]. Each Hydrogen atom contains one proton ($p^+$), and one electron ($e^-$). Each Helium atom contains $2p^+$, $2e^-$ and two neutrons ($n^0$). These are summarized below:

*Hydrogen* $\rightarrow 1p^+ + 1e^-$
*Helium* $\rightarrow 2p^+ + 2e^- + 2n^0$
*Heavy elements* $\rightarrow xp^+ + xe^- + yn^0$, where $x, y > 2$.

The x and y values must be known to perform an exact calculation. Since the heavy elements represent only 2% of the observable universe, one option is to use average x and y values representing all the heavy elements. However, in this study, in order to get an accurate picture, we will perform the exact calculation for all the known elements contained in the universe, including the huge range of heavy elements. The data table in the Supplementary Material shows the percentage weight abundance of all the elements in the observable universe. The key aspect here is to realize that the percentage weight abundance must be converted into percentage number of atoms abundance, if one needs to estimate the total number of elementary particles in the universe and their information content. This is done using the molar mass values (units of moles/g) for each element, to work out the moles (number of atoms) of each kind. This calculation shows that the percentage weight abundance of elements in the universe is very different to the percentage of atoms abundance. For example, % weight values for Hydrogen and Helium are 75% and 23%, respectively, while their % numbers of atoms are 92.68% and 7.15%, respectively (see the Supplementary Material for full list of values).

Taking into account the internal structure of the atoms by using the exact number of $p^+$, $e^-$ and $n^0$ in each kind of atom, and using the % abundance of atoms determined, we can work out the total number of $p^+$, $e^-$ and $n^0$ corresponding to each kind of atom. Summing up all $p^+$, $e^-$ and $n^0$ we obtain: 108.27 electrons, 108.27 protons, and 15.6 neutrons, where the fractional values reflect the probabilistic nature of the estimates. The total statistical number of atomic constituents representative for the matter in the observable universe is then a total of 232.14 particles ($p^+$, $e^-$ and $n^0$). This allows us to estimate the probabilities of observing a $p^+$, $e^-$ and $n^0$ in the observable universe as: $P_{p+} = P_{e-} = 108.27 / 232.14 = 0.466$ and $P_{n0} = 15.6 / 232.14 = 0.067$. We can now regard this baryonic distribution of matter in the observable universe as a set of three possible states / events, $\{p^+, e^-, n^0\}$, with the occurrence probabilities $\{P_{p+}, P_{e-}, P_{n0}\} = \{0.466, 0.466, 0.067\}$. Shannon's classical information theory gives us an elegant mechanism to estimate the bits information content per event. In the case of a three-state system, the maximum information ($I$) encoded per event is $I = log_2 3 = 1.585$ bits, corresponding to the case when the three events / states are equally probable. This is not the case here, so the average bits information content per event, meaning per any element of the set $\{p^+, e^-, n^0\}$, is given by:

$$H_{pen} = -\sum_{j=1}^{n} P_j \cdot \log_2 P_j = -\left(P_{p+} \log_2 P_{p+} + P_{e-} \log_2 P_{e-} + P_{n0} \log_2 P_{n0}\right) \quad (6)$$

Introducing the numerical values in (6), we obtain 1.288 bits of information encoded per elementary particle. The meaning of this result is that the most effective way that information can be compressed into a $p^+$, $e^-$ or $n^0$, or the average amount of information content stored in each of these particles, is 1.288 bits per particle.

To find out the total information stored in all the matter particles in the universe, one needs to estimate the total number of protons, electrons and neutrons in the observable universe, and to multiply that number by the bits information per elementary particle, i.e. 1.288.

Let us remember that the information content estimated above is in fact inaccurate as the protons and neutrons are not elementary particles and they have an internal structure containing three quarks each. According to the standard model, a proton contains two up quarks and one down quark, and a neutron contains one up quark and two down quarks. Hence, we can estimate the number of up / down quarks per each kind of atom, and using the % abundance of atoms, we can estimate how many up and down quarks are in a sample of 100 atoms. Summing up we obtain a total of 216.55 up quarks and 108.27 down quarks contained in protons, and 15.6 up quarks and 31.21 down quarks contained in neutrons. This results in a total of 232.15 up quarks and 139.48 down quarks. Adding up the total number of electrons (108.27) to the total number of up quarks and down quarks, we obtain 479.9, which is the total number of elementary particles contained in a statistical sample of 100 random atoms. Using the acronyms $uq$ for up quarks and $dq$ for down quarks, we can now regard, again, these elementary particles as a set of three possible states / events, where the three possible states are $\{uq, dq, e^-\}$, and their occurrence probabilities are $\{P_{uq}, P_{dq}, P_{e-}\}$. The occurrence probabilities are obtained as $P_{uq} = 232.15 / 479.9$, $P_{dq} = 139.48 / 479.9$ and $P_{e-} = 108.27 / 479.9$, resulting in $\{P_{uq}, P_{dq}, P_{e-}\} = \{0.483, 0.29, 0.225\}$.

The Shannon's information entropy formula for this set of events and probabilities is:

$$H_{uqdqe} = -\sum_{j=1}^{n} P_j \cdot \log_2 P_j = -\left(P_{uq} \log_2 P_{uq} + P_{dq} \log_2 P_{dq} + P_{e-} \log_2 P_{e-}\right) \quad (7)$$

Introducing the numerical values in (7) we obtain 1.509 bits of information encoded per elementary particle. If the information content per elementary particle in the observable universe is known, the total information stored in the baryonic matter in the universe could be worked out if one knows how many particles, $N_b$, are in the universe.

## 3. The total number of particles in the universe

The total number of particles in the universe, protons more precisely, is known as the Eddington number, $N_{Edd}$, and it was first calculated in 1938 [32]. Eddington estimated that the universe contains $N_{Edd} = 136 \times 2^{256}$, or about $1.57 \times 10^{79}$ protons.

Here we provide an analytical formula for the total number of baryons in the universe, which allows a more detailed estimation of the number of particles in the universe. Using Friedmann equation [33] for a homogeneous, isotropic and flat universe, we determine the well-known critical density, $\rho_c$, of the universe as:

$$\rho_c = \frac{3H^2}{8\pi G} \quad (8)$$

where $H = 2.2 \times 10^{-18}$ $1/s$ is the Hubble parameter today and $G = 6.674 \times 10^{-11}$ $m^3/Kg \cdot s^2$ is the gravitational constant. The critical density is then $\rho_c = 9.6 \times 10^{-27}$ $Kg/m^3$. The fraction of the energy density stored in baryons is $\Omega_b = 0.0485$ [34], which allows us to estimate the baryons density as $\rho_b = \Omega_b \cdot \rho_c$. The radius of the observable universe is estimated to be about 46.5 billion light-years [35], or $L_u = 44 \times 10^{25}$ $m$. The volume of the observable universe is then $V_u = 4/3 \cdot \pi \cdot L^3 = 4 \times 10^{80}$ $m^3$. The total mass of ordinary matter in the universe, or the mass of all baryons in the observable universe, $M_b$, can be calculated as:

$$M_b = \rho_b \cdot V_u = \frac{H^2 \Omega_b L_u^3}{2G} \quad (9)$$

Considering the probabilities of occurring of each $p^+$, $e^-$ and $n^0$, $\{P_{p+}, P_{e-}, P_{n0}\} = \{0.466, 0.466, 0.067\}$, the effective mass is defined:

$$m^b_{eff} = P_{e-}m_{e-} + P_{p+}m_{p+} + P_{n0}m_{n0} \quad (10)$$

The $m_{e-} = 9.109 \times 10^{-31}$ $kg$, $m_{p+} = 1.672 \times 10^{-27}$ $kg$ and $m_{n0} = 1.675 \times 10^{-27}$ $kg$ are the rest masses of the electron, proton and neutron, respectively. The number of baryons in the universe is then obtained by dividing the mass of all the baryons in the universe by the effective mass:

$$N_b = \frac{H^2 \Omega_b L_u^3}{2G \cdot m^b_{eff}} \quad (11)$$

Performing the numerical calculations, we obtain $N_b = 1.93 \times 10^{80}$ baryons in the universe (more precisely the total number of particles in the universe as it includes electrons, which are not baryons). This can be used to determine the exact number of electrons ($N_{e-}$), protons ($N_{p+}$) and neutrons ($N_{no}$), as:

$$N_{e-} = P_{e-}N_b \; ; N_{p+} = P_{p+}N_b \; ; N_{n0} = P_{n0}N_b \quad (12)$$

The numerical values are $N_{e-} = N_{p+} = 9.027 \times 10^{79}$, and $N_{n0} = 1.298 \times 10^{79}$. Since each neutron and proton contains 3 quarks, the total number of particles in the observable universe is: $N_{tot} = N_{e-} + 3(N_{p+} + N_{no})$, or:

$$N_{tot} = \frac{H^2 \Omega_b L_u^3}{2G \cdot m_{eff}^b} \left( P_{e-} + 3(P_{p+} + P_{n0}) \right) \quad (13)$$

The calculations indicate that the observable universe contains $N_{tot} = 4 \times 10^{80}$ particles, which is in good agreement with Eddington's number giving only the number of protons.

## 4. Total information content and conclusions

Since we calculated in section 2 that each particle stores $H_{uqdqe} = 1.509$ bits of information, the total information content of the baryonic matter in the observable universe is $N_{bits} = N_{tot} \times H_{uqdqe} = 6.036 \times 10^{80}$ bits. This value is lower than the previous estimates [6,8,22,24,28-30]. However, our calculation refers only to the information stored in particles such as electrons, protons, neutrons and their constituent quarks. The previous estimates refer to the total information content of the universe. Moreover, our calculation takes also a different approach, using the information theory, which gives the most effective information compression. The discrepancy in the information content values could therefore be explained by the fact that our estimate is for the most effective compression mechanism, according to the information theory. Other possible reasons for the information content discrepancy could perhaps suggest that the universe most likely contains more information, which is stored in other elementary particles, or other media, not accounted for in this study. For example, we have considered all bosons to be force / interaction particles responsible for the transfer of information, rather than store of information. We also ignored all the anti-particles, as well as all neutrinos. Although there is scope for further refinements of this work, here it is the first time that the information theory has been used to estimate the information content of the observable universe, and this formalism most certainly would stimulate future studies. Moreover, the current approach offers a unique tool of estimating the information content per elementary particle, which is very useful for designing practical experiments to test these predictions.

## 5. Supplementary Material

The Supplementary Material contains the known percentage weight abundance of all the elements in the observable universe [36] and their estimated abundance by the number of atoms. The Supplementary Material also contains the number of electrons, protons and neutrons per each element, which allows estimating their occurrence probabilities.

## Acknowledgments

The author acknowledges the financial support received to undertake this research from the School of Mathematics and Physics, University of Portsmouth.

## Data Availability

The numerical data associated with this work is available in the Supplementary Material and the detailed calculations are available by request from the author.